\documentclass{aa}  
\usepackage{graphicx}
\usepackage{amsmath}	% Advanced maths commands
\usepackage{amssymb}	% Extra maths symbols
\usepackage{relsize}
\usepackage{txfonts}
\begin{document} 

   \title{Counting stars from the integrated spectra of galaxies}
   \subtitle{}

   \author{
I. Mart\'in-Navarro \inst{1,2} \& A. Vazdekis \inst{1,2} 
          }
   \institute{
Instituto de Astrof\'{\i}sica de Canarias,c/ V\'{\i}a L\'actea s/n, E38205 - La Laguna, Tenerife, Spain\\
\email{imartin@iac.es, vazdekis@iac.es}
\and
Departamento de Astrof\'isica, Universidad de La Laguna, E-38205 La Laguna, Tenerife, Spain
}

   \date{Received, accepted}
   
   \titlerunning{Counting stars from integrated spectra}
\authorrunning{Mart\'in-Navarro}  
 
  \abstract
   {
    Over the last decades, evolutionary population synthesis models have powered an unmatched leap forward in our understanding of galaxies. From dating the age of the first galaxies in the Universe to detailed measurements of the chemical composition of nearby galaxies, the success of this approach built upon simple stellar population (SSP) spectro-photometric models is unquestionable. However, the internal constraints inherent to the construction of SSP models may hinder our ability to analyze the integrated spectra of galaxies in situations where the SSP assumption does not sufficiently hold. Thus, here we revisit the possibilities of fitting galaxy spectra as a linear combination of stellar templates without assuming any a priori knowledge on stellar evolution. We showcase the sensitivity of this alternative approach to changes in the stellar population properties, in particular the direct connection to variations in the stellar initial mass function, as well as its advantages when dealing with non-canonical integrated populations and semi-resolved observations. Furthermore, our analysis demonstrates that the absorption spectra of galaxies can be used to independently constrain stellar evolution theory beyond the limited conditions of the solar neighborhood.
   }


   \maketitle
%
%-------------------------------------------------------------------

\section{Introduction} \label{sec:intro}

The stellar absorption spectrum of a galaxy is the linear combination of the flux emitted by its individual stars. Although this statement may sound obvious, it has immediate implications since it enables us to anchor and interpret observations of distant galaxies based on the well-calibrated properties and thoroughly tested physics of nearby stars. Arguably, the most direct way to analyze the integrated spectrum of a galaxy is to compare it to a set of stellar spectra, also known as empirical modelling. Finding a linear combination of stars that can reproduce an observed spectrum was early recognized as a viable way forward \citep{Ohman34,Whipple35} and since then, variations of this method (including the use of star cluster spectra as templates) led to substantial advances in our description of the stellar population content of galaxies \citep[e.g.][]{Spinrad71,Faber72,Pickles,Bica86,Bica88,Schmidt91}. The main difficulty that these initial attempts found was, however, the uniqueness of the recovered solution \citep[e.g.][]{Pelat98} and thus the consistency of the inferred properties across different wavelength ranges \citep[e.g.][]{Eftekhari22}.

Instead of freely combining individual stellar spectra, knowledge on stellar evolution theory can also be assumed in order to model the spectrum of a galaxy. This approach, pioneered by the seminal work of Beatrice Tinsley \citep{Tinsley68,Tinsley72,Tinsley76,Tinsley80}, sets fixed but physically-motivated internal constraints on both the properties \citep[through theoretical isochrones, e.g.,][]{Padova94,basti1,Choi16} and the relative number of stars \citep[i.e. the initial mass function, IMF, e.g.,][]{kroupa,Chabrier} that are required to generate the model spectrum of a stellar population. This so-called evolutionary population synthesis technique predicts the absorption spectrum of a simple stellar population (SSP) given its age and chemical composition\footnote{Nowadays, the term SSP usually refers to the actual spectral model of a stellar population.}. Over the last decades, SSP model predictions have reached an exquisite level of refinement \citep[e.g.,][]{Leitherer99,bc03,TMB:03,Schiavon07,miles,Conroy12}, which in turn has revolutionized our view of galaxy formation and evolution \citep[e.g.,][]{Worthey92,vazdekis:97,Gallazzi05,Thomas05,Kuntschner06,vandokkum,labarbera,McDermid15}.

In parallel to the development of SSP models, the mathematical tools to analyze the integrated spectra of galaxies have also experienced a significant leap forward. In particular, spectral fitting algorithms able to model an observed spectrum as a linear combinations of SSPs are now widely available \citep[e.g.][]{ppxf,CF05,Ocvirk06a,Vespa,Koleva09,pipe3d,Carnall,prospector}, allowing also the analysis of populations with complex and extended star formation histories. 

The success of evolutionary population synthesis models is, therefore, irrefutable. However, the hard-coded assumptions inherent to this approach are not free from biases and potential flaws. For example, fundamental ingredients in the calculation of theoretical isochrones \citep[opacities, mass losses, mixing length, convection etc. see e.g.,][]{Maraston05,Conroy09} are generally weakly constrained or approximated. The decades-long debate about the morphology of the horizontal branch is a paradigmatic example of the current limitations of stellar evolution theory \citep[e.g.][]{Sandage67,Lee94,Dotter07}. In addition, non-canonical evolutionary pathways such as the contribution of binary stars or extreme mass-loss phases can also have an important contribution to the observed spectrum yet they remain poorly incorporated into SSP model predictions \citep[e.g.][]{Renzini88,Greggio90,Eldridge17}.

Two recent developments have further highlighted some of the limitations of evolutionary population synthesis models. First, the apparent non-universality of the IMF \citep[e.g.][]{ferreras,Spiniello2013, MN15a, Parikh} casts doubts on the flexibility of SSP models to account for complex IMF variations \citep{Conroy17}, particularly when dealing with young stellar populations \citep{MN24}. Finally, the advent of a new generation of spectroscopic facilities such as the Local Volume Mapper \citep{LVM} and most importantly the European Extremely Large Telescope \citep{ELT} and the US Extremely Large Telescope Program \citep{uselt} will observe the nearby Universe in a semi-resolved manner, where the basic assumption behind SSP models of a fully sampled IMF does not longer hold.

In this letter, we revisit the possibilities of a more agnostic approach to stellar population inference, fitting integrated absorption spectra of galaxies as a linear combination of stellar spectra. Equipped with state-of-the-art spectral fitting algorithms and stellar libraries, we demonstrate some of the unique advantages of this approach as well as its potential limitations. The outline of this work is as follows: in \S~\ref{sec:linear} we describe the fitting scheme and in \S~\ref{sec:tests} we present a series of consistency tests. Finally, in \S~\ref{sec:discu} we discuss our results and point towards some interesting prospects.

\section{Fitting scheme} \label{sec:linear}

We based our analysis on the Penalized PiXel-Fitting (pPXF) algorithm \citep{ppxf}. In short, pPXF was designed to fit the integrated spectra of galaxies as a linear combination of SSP model predictions. Thus, the basic output of pPXF is a weights matrix defining the relative contribution of each SSP model to the observed spectrum. Moreover, as described in detail in \citep{Cappellari17}, pPXF is able to retrieve robust stellar population measurements by regularizing the recovered weights in up to three dimensions. Then, pPXF is usually paired with a set of SSP models covering a range of ages, metallicities, and [$\alpha$/Fe] to derive star formation histories and chemical enrichment patterns. 

From the point of view of individual stars, a library of stellar spectra is also characterized by three main parameters, namely $\log g$, $T_{eff}$ and [Fe/H]. With this idea in mind, we paired pPXF with the MILES stellar library presented in \citet{Pat06}. However, MILES is an empirical stellar library and thus stars are not homogeneously distributed across the whole $\log g$-$T_{eff}$-[Fe/H] parameter space. Therefore, to maximize the capabilities of pPXF, we used the roughly 1,000 stars from the MILES library and the interpolation scheme described in \citet{vaz03} and updated in \citet{miles} to generate a regular grid of stellar spectra. It is worth highlighting that, according to stellar evolution theory, stars with different masses (and thus different $\log g$ and $T_{eff}$) have different luminosities. However, we intentionally did not want to include any additional information during the fitting process. Hence, we neglected the expected differences in luminosity and the flux of each stellar template was normalized to the same value, regardless of its mass. The importance of this decision will become obvious in the next section.

\section{Results} \label{sec:tests}

\subsection{Fitting SSP models}

\begin{figure}
    \centering
    \includegraphics[width=8.4cm]{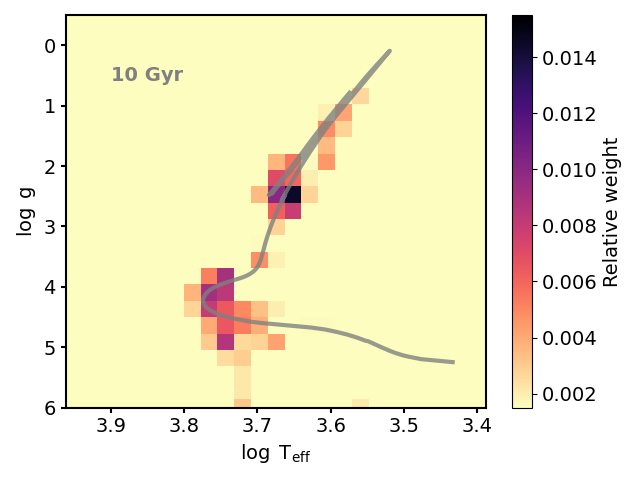}
    \caption{Best-fitting solution for an SSP model. Each pixel in the image corresponds to the relative flux contribution of stars with different $\log g$ and $T_{eff}$ that better reproduce a MILES SSP model of 10 Gyr and solar metallicity. On top of the weights, the grey line indicates a theoretical isochrone with the same age and metallicity used to build up the MILES SSP model.}
    \label{fig:ssp}
 \end{figure}

An immediate way to assess the reliability of our approach is to fit, using individual stars, SSP model predictions constructed with the very same stellar spectra. In Fig.~\ref{fig:ssp} we show the flux weights distribution recovered when fitting a MILES model of 10 Gyr and solar metallicity with the MILES stellar library. Over-plotted on top of the weights distribution, we also show a \citet[][scaled solar]{basti1} theoretical isochrone for a 10 Gyr and solar metallicity population as well used to generate the MILES SSP model.  

It is clear from Fig.~\ref{fig:ssp} that the flux weights recovered by pPXF are not randomly distributed but closely follow the isochrone track. This result is particularly noteworthy because no stellar evolution knowledge has been implemented in the code and the inversion problem is highly degenerate \citep{Schmidt91,Eftekhari22}, even more than in the case of SSP models \citep[e.g.][]{Worthey94}, since for a single SSP model pPXF has to explore a parameter space as large as it would be required to measure the whole star formation history of a galaxy. Yet, pPXF fed with the MILES stellar library is able to retrieve a physically-meaningful solution. Note that the value of the recovered weights along the isochrone depend on two main factors: the intrinsic luminosity of stars with that particular mass and the number of them (i.e. the IMF). Therefore, the flux weights are dominated by the contribution of luminous giant stars (the prominence of the \textit{red clump} is particularly evident) and by the relatively bright and heavily-populated main sequence turnoff.

\begin{figure*}
    \centering
    \includegraphics[width=6.1cm]{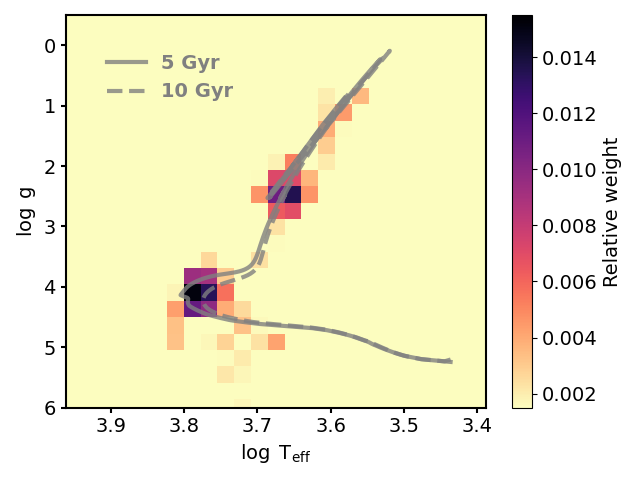}
    \includegraphics[width=6.1cm]{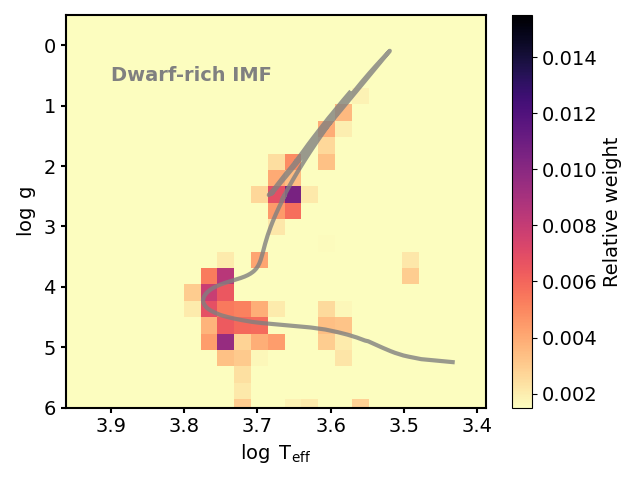}
    \includegraphics[width=6.cm]{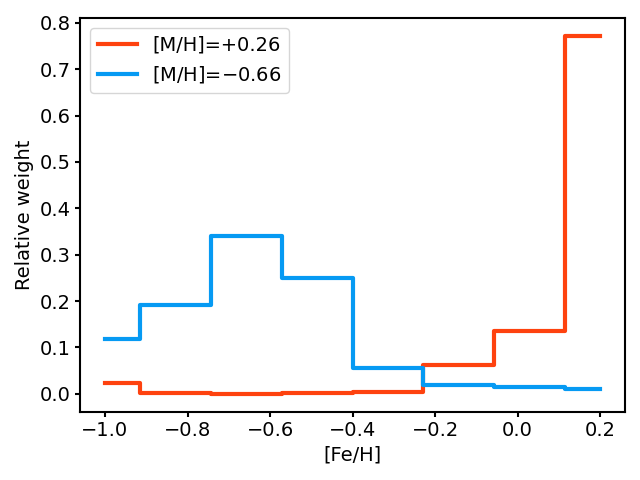}

    \caption{Sensitivity to changes in the stellar population properties. (Left) Weights distribution retrieved after fitting the MILES SSP model prediction of a population with solar metallicity and an age of 5 Gyr, with two isochrones of 5 and 10 Gyr over-plotted for comparison. (Middle) Similar weights distribution but for an old (10 Gyr) and solar metallicity population with a relative excess of low-mass stars (i.e. a bottom-heavy IMF). In this case, the isochrone corresponds to a 10 Gyr population. (Right) Recovered metallicity weights after fitting two MILES SSP models, in red and blue for a metal-rich and a metal-poor population, respectively.}
    \label{fig:tests}
 \end{figure*}

To expand this series of consistency tests, we show in Fig.~\ref{fig:tests} the result of fitting SSP models with different ages, IMF slopes and metallicities. On the left panel, the weights distribution of a young 5 Gyr population and solar metallicity is show. While there is a small yet noticeable difference in the position of the main sequence turn-off, the most evident change with respect to Fig.~\ref{fig:ssp} is the relatively larger weights given to these turn-off stars when fitting the 5 Gyr population, which simply reflects the presence of hotter and more massive (/luminous) stars. Interestingly, changes in the IMF slope are also propagated to the recovered weights distribution as shown in the middle panel of Fig.~\ref{fig:tests}. In this case, we fit again a SSP model of 10 Gyr and solar metallicity but with a bottom-heavy (i.e. dwarf-rich) IMF. There are two key differences between this middle panel and the results shown in Fig.~\ref{fig:ssp}. First, the relative weight of dwarf-to-giants is higher in this case, as expected from the change in the IMF slope. Second, pPXF tends to give weights to stars with even lower stellar masses (i.e. $\log g \gtrsim 4.5$ and  $\log T_{eff} \lesssim 3.6$). Thus, in an idealized scenario, the proposed setup can also be sensitive, in a non-parametric way, to changes in the IMF. This, however, would require a physically-motivated normalization for each stellar template. Finally, the right panel of Fig.~\ref{fig:tests} shows the recovered metallicity distribution for two SSP models: in red for a metal rich population ([M/H]=0.26) and in blue for a metal-poor one ([M/H]=-0.66). Again, changes in the metallicity of the underlying stellar population are systematically recovered.

 \subsection{Non-canonical stellar populations}

An obvious advantage of our fitting scheme is the possibility of dealing with non-canonical stellar populations. To exemplify this, we briefly take on the results presented in \citet{Nuria22}, where the authors discuss two scenarios to explain the line-strength ratios of UV indices in the spectrum of the massive galaxy NGC\,1277. On the one hand, the  measured line-strength indices can be explained by the presence of some residual star formation and thus young, massive stars on top of an underlying old stellar population. On the other, the observed line-strength ratios can also be interpreted as a combination of the same old stellar population plus the contribution of an additional similarly old population with a peculiarly extreme horizontal branch. 

\begin{figure}
    \centering
    \includegraphics[width=8.4cm]{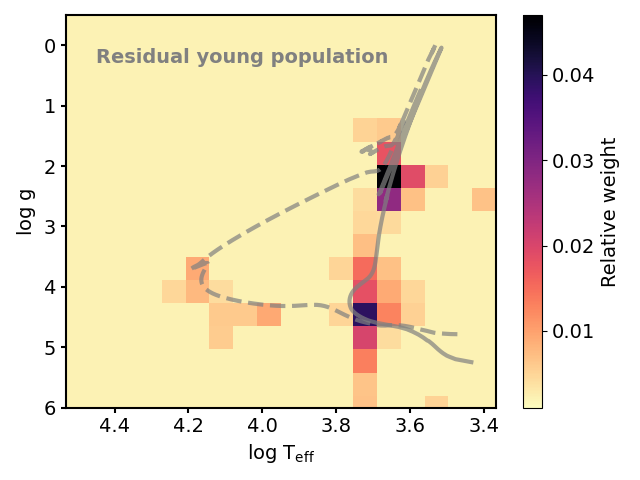}
    \includegraphics[width=8.4cm]{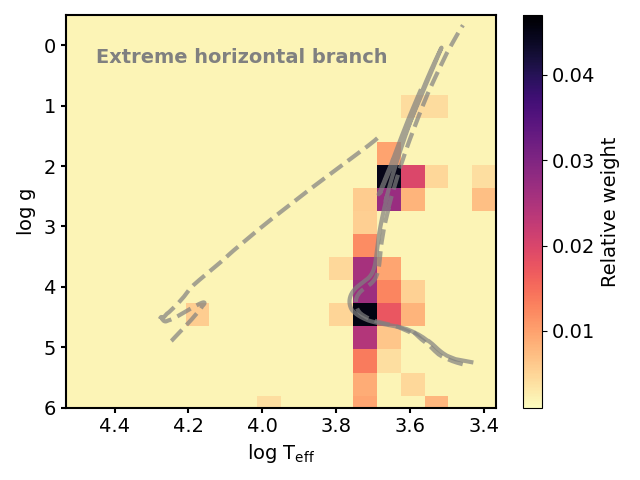}

    \caption{Young populations vs extreme horizontal branch stars. (Top) Weights distribution measured for a synthetic spectrum combining a dominant old (13 Gyr) population and an 0.5\% contribution of an SSP with $\sim0.1$ Gyr. Solid and dashed grey lines indicate the isochrones of the old and young populations, respectively. The recovered weights follow the tracks defined by both isochrones. (Bottom) Similar weights distribution but for the combination of the same old population plus a 40\% of an SSP with an extreme horizontal branch. Again, solid and dashed grey lines indicate the theoretical isochrones used to build the models.}
    \label{fig:hb_young}
 \end{figure}

In particular, \citet{Nuria22} invoke a combination of a $\sim$13 Gyr population plus a 0.5\% contribution of $\sim0.1$ Gyr for the first scenario, or the same old population combined with a 40\% of another SSP characterized by an extreme horizontal branch. Using these values, we generated two synthetic spectra corresponding to each hypothesis and analyzed them using our fitting approach. The result of this test is shown in \ref{fig:hb_young}.

The top panel of \ref{fig:hb_young} shows the recovered weights for the case of a residual young population on top of the dominant old component. Overplotted to these weights we also show the isochrones assumed to generate the two populations. In addition to the main 13 Gyr population that dominates the weights distribution, stars around the main sequence turnoff of the secondary young population are also clearly recovered by the fitting scheme. On the bottom panel, corresponding to the second scenario with an extreme horizontal branch population, the measured weights distribution is clearly different, lacking of the characteristic turnoff stars shown in the upper panel. 

From this test, summarized in Fig.~\ref{fig:hb_young}, it is evident the strength of this approach to fit, in a non-parametric way, the absorption spectra of stellar populations that depart from the internal assumptions of SSP models.

\subsection{Fitting observed galaxy spectra}

The results above are idealized scenarios to demonstrate some of the potential applications of a physics-free fit to the integrated spectra of galaxies. To test the feasibility of this approach to deal with real data, we analyzed the Sloan Digital Sky Survey \citep{DR6} stacked spectra of \citet{labarbera}. In short, they consist of sixteen high signal-to-noise stacked spectra of nearby early-type galaxies (ETGs) covering a range of stellar velocity dispersions. These spectra have been previously studied in detail \citep[see e.g.][]{ferreras,labarbera,Rosani18,Pernet24} and thus we can use them to assess the robustness of our approach. 

\begin{figure}
    \centering
    \includegraphics[width=8.6cm]{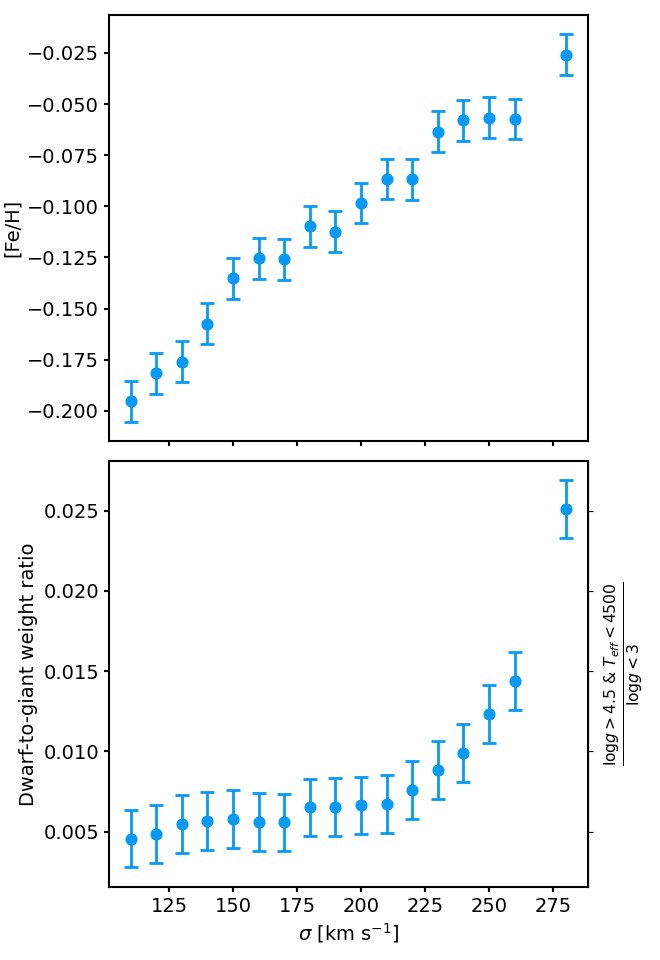}

    \caption{Fits to SDSS data. (Top) Measured iron metallicity values as a function of stellar velocity dispersion for the SDSS stacked spectra. We recover the well-known relation between the metallicity of galaxies and their velocity dispersions based on the linear combination of MILES stars. Note that metallicity refers here to [Fe/H] as it is based on the metallicity of these individual stars. (Bottom) Ratio between the weight given to low-mass and high-mass stars by our fitting approach. This weight ratio is a proxy for variations in the IMF and suggests a relative excess of low-mass stars in the most massive ETGs.}
    \label{fig:sdss}
 \end{figure}

The top panel of Fig.~\ref{fig:sdss} shows the measured iron metallicity of the SDSS stacked spectra based on the combination of MILES stars as a function of stellar velocity dispersion $\sigma$. As anticipated by the right panel in Fig.~\ref{fig:tests}, the proposed approach is sensitive to changes in the metallicity content of the absorption spectra, recovering in this case the well known metallicity-$\sigma$ relation. It is worth noting that our measurements are based on the iron metallicity ([Fe/H]) of the individual MILES stars. In this regard, the recovered trend coincides with the line-strength analysis of \citet{labarbera}, who found a variation in [Fe/H] from -0.17 to -0.04 between the low-$\sigma$ and high-$\sigma$ SDSS stacked spectra, once corrected from the measured [Mg/Fe] abundance ratio. 

Inspired by the comparison between Fig.~\ref{fig:ssp} and the middle panel in Fig.~\ref{fig:tests}, we also assessed the possibility of measuring the effect of a variable IMF. For the SDSS sample, detailed line-strength (SSP-based) measurements have revealed a systematic variation in the IMF slope, with galaxies with higher $\sigma$ exhibiting an enhanced fraction of low-mass stars \citep{ferreras,labarbera,MN19,denBrok24}. Our approach is directly connected to the IMF of the analyzed stellar population since the measured weights are effectively mapping number of stars as a function of their mass, modulated by their luminosity. Moreover, the proposed approach probes the IMF in a non-parametric way, allowing in principle for any kind of underlying change in the IMF. 

Capitalizing on this key feature, the bottom panel of Fig.~\ref{fig:sdss} shows the ratio between the flux weight given to low-mass M-dwarf stars ($\log g > 4.5$ and T$_{eff}<4500$) and the flux weight given to giant stars ($\log g < 3$). Because of the rather old ages of the sample, this metric approximates the dwarf-to-giant ratio traditionally measured in ETGs to quantify changes in the IMF slope. From the bottom panel of Fig.~\ref{fig:sdss} it is clear that our approach also points towards a systematic change in the IMF: the weights given to low-mass stars in order to fit the observed spectra become progressively more important for more massive galaxies. This finding is in remarkable agreement with the expectations based on alternative stellar population analyses.

\section{Discussion and prospects} \label{sec:discu}

Interpreting the integrated spectra of galaxies with as few model assumptions as possible is the ultimate goal of any stellar population analysis. In this context, the development of evolutionary population synthesis models has been a necessary and extremely successful stepping stone. Yet, we have demonstrated here that it is now possible to further distill down the modelling process, reducing it to its most basic components: the spectra of individual stars. The fact that the measured weights shown in e.g. Fig.~\ref{fig:ssp} are not sparsely distributed but follow the track of theoretical isochrones is a striking result. It shows that there is physically-meaningful information about stellar evolution theory and about the number distribution of stars encoded in the integrated spectra of galaxies.

Two immediate applications follow these findings. First, it is possible to use integrated spectra, decomposed into the flux of individual stars, to constrain the physical ingredients that go into the calculation process of evolutionary population synthesis models, and more generally, to improve our knowledge of stellar evolution theory. Second, it provides a more direct access to studying the IMF of unresolved stellar populations.  In particular, the rigidity of SSP models has so far severely hampered our ability to compare dynamical and stellar population-based IMF measurements \citep[e.g.][]{Oldham,Adriano21}. The model-independent approach sketched here (see also \citealt{Dries16}) presents a unique opportunity to model in a self-consistent way the properties and dynamics of individual stars beyond the Milky Way \citep[e.g.][]{Glenn08,Ling18b,Vasiliev20}.

It is also worth mentioning that, although the results above are already promising, there are several aspects of the fitting scheme that can be significantly improved. For example, while the way that the regularization of the solution is implemented in pPXF is ideally suited to work with SSP models where rather smooth changes are expected in the star formation history and chemical enrichment (and thus on the recovered weights), it does not necessarily apply to the weight distribution of individual stars since the expected smoothness of the solution is high along the isochrone but weak on the orthogonal direction. Moreover, metallicity, luminosity, $\log g$, and T$_{eff}$ are not fully independent quantities and thus the degree of freedom of the inversion problem tackled in this work, and consequently the associated uncertainties, have been artificially amplified. Finally, translating weights into physical quantities such as ages or chemical compositions would require an additional modelling layer, in a similar same way to resolved stellar populations analyses \citep[e.g.][]{Gallart05}. Ultimately, the proposed approach will require dedicated tools and templates to fully maximize its potential applications.

With all this in mind, could a physics-free approach replace the use of SSP models? In the foreseeable future, SSP model predictions will likely remain as a fundamental tool for stellar population analyses. In many situations, the complexity and intrinsic degeneracies of the inversion problem, amplified by the noisy nature of real data, can only be partially mitigated through the internal constraints of evolutionary population synthesis models. Ultimately, our ability to retrieve information about the stellar population properties of galaxies depends on the (limited) amount of information encoded in their absorption spectra \citep[e.g.][]{Ferreras23}. 

However, as demonstrated here, there are several advantages to fitting the absorption spectra of galaxies using stellar spectra and both approaches can actually complement each other. Moreover, there are some important questions that can only be tackled outside of the rigid limits of SSP models. With larger and better stellar libraries \citep{MILES21,Mastar,Knowles21}, the advent of a new generation of observational facilities, and the current explosion of computational capabilities, it is possible to rethink our approach to stellar population models, aspiring to gain access to new information about the star formation processes in galaxies.

\begin{acknowledgement}

We would like to thank the comments and suggestions from the referee which helped improving the manuscript. IMN would like to thank Bron Reichardt-Chu, Dimitri Gadotti, Sebastián Sánchez, Glenn van de Ven and the TRACES group for the discussions that sparked some of the ideas motivating this work. We acknowledge support from grant PID2022-140869NB-I00 from the Spanish Ministry of Science and Innovation.

\end{acknowledgement}

\bibliographystyle{aa}  % style aa.bst
% \bibliography{starpaint} % your references Yourfile.bib 

\end{document}